\documentclass[aps,preprint]{revtex4}
\usepackage{amssymb,epsf}
\usepackage{amsmath}
\usepackage{graphicx}
\begin{document}

\title{Taub-NUT Black Holes in Third order Lovelock Gravity}
\author{S. H. Hendi$^{1,3}$\footnote{email address: hendi@mail.yu.ac.ir}
and M. H. Dehghani$^{2,3}$ \footnote{email address:
mhd@shirazu.ac.ir}}
\affiliation{$^1$Physics Department, College of Sciences, Yasouj University, Yasouj
75914, Iran\\
$^2$Physics Department and Biruni Observatory, College of Sciences, Shiraz
University, Shiraz 71454, Iran\\
$^3$Research Institute for Astrophysics and Astronomy of Maragha (RIAAM),
Maragha 55134, Iran}

\begin{abstract}
We consider the existence of Taub-NUT solutions in third order Lovelock gravity
with cosmological constant, and obtain the general form of these
solutions in eight dimensions. We find that, as in the case of Gauss-Bonnet gravity
and in contrast with the Taub-NUT solutions of Einstein gravity, the metric function depends on the specific
form of the base factors on which one constructs the circle fibration. Thus, one may say that
the independence of the NUT solutions on the geometry of the base space is not a
robust feature of all generally covariant theories of gravity and is peculiar to Einstein
gravity. We find that when Einstein gravity admits non-extremal NUT solutions
with no curvature singularity at $r=N$, then there exists a non-extremal NUT
solution in third order Lovelock gravity. In $8$-dimensional spacetime,
this happens when the metric of the base space is chosen to be $\Bbb{CP}^{3}$.
Indeed, third order Lovelock gravity does not admit non-extreme NUT solutions with any other
base space. This is another property which is peculiar to Einstein gravity.
We also find that the third order Lovelock gravity admits extremal NUT solution
when the base space is $T^{2}\times T^{2}\times T^{2}$ or $S^{2}\times T^{2}\times T^{2}$.
We have extended these observations to two conjectures about the existence of NUT solutions
in Lovelock gravity in any even-dimensional spacetime.
\end{abstract}

\maketitle
\section{Introduction}

The question as to why the Planck and electroweak scales differ by so many
orders of magnitude remains mysterious. In recent years, attempts have been
made to address this hierarchy issue within the context of theories with
extra spatial dimensions. In higher-dimensional spacetimes even with the
assumption of Einstein -- that the left-hand side of the field equations is
the most general symmetric conserved tensor containing no more than two
derivatives of the metric -- the field equations need to be generalized.
This generalization has been done by Lovelock \cite{Lov}, and he found a
second rank symmetric conserved tensor in $d$ dimensions which contains upto
second order derivative of the metric. Other higher curvature gravities
which have higher derivative terms of the metric, e.g., terms with quartic
derivatives, have serious problems with the presence of tachyons and ghosts
as well as with  perturbative unitarity, while the Lovelock gravity is free
of these problems \cite{Zw}.

Many authors have considered the possibility of higher curvature terms in
the field equations and how their existence would modify the predictions
about the gravitating system. Here, we are interested in the properties of
the black holes, and we want to know which properties of the black holes are
peculiar to Einstein gravity, and which are robust features of all
generally covariant theories of gravity. This fact provide a strong motivation for
considering new exact solutions of Lovelock gravity. We show that some properties of
NUT solutions are peculiar to Einstein gravity and not robust feature of all
generally covariant theories of gravity. Although the nonlinearity of the
field equations causes to have a few exact black hole solutions in Lovelock
gravity, there are many papers on this subject \cite{All GB
Third,DehMan,DehHen}. In this letter we
introduce Taub-NUT metrics in third order Lovelock gravity, and investigate
which properties of these kinds of solutions will be modified by considering
higher curvature terms in the field equations.

The original four-dimensional
solution \cite{TaubNUT} is only locally asymptotic flat. The spacetime has
as a boundary at infinity a twisted $S^{1}$ bundle over $S^{2}$, instead of
simply being $S^{1}\times S^{2}$. There are known extensions of the Taub-NUT
solutions to the case when a cosmological constant is present. In this case
the asymptotic structure is only locally de Sitter (for positive
cosmological constant) or anti-de Sitter (for negative cosmological
constant) and the solutions are referred to as Taub-NUT-(A)dS metrics. In
general, the Killing vector that corresponds to the coordinate that
parameterizes the fibre $S^{1}$ can have a zero-dimensional fixed point set
(called a NUT solution) or a two-dimensional fixed point set (referred to as
a `bolt' solution). Generalizations to higher dimensions follow closely the
four-dimensional case \cite{BaisPA,Awad,Lorenzo}. Also, these kinds
of solutions have been generalized in the presence of electromagnetic field
and their thermodynamics have been investigated \cite{Mann12,DehKhod}. It is
therefore natural to suppose that the generalization of these solutions to
the case of Lovelock gravity, which is the low energy limit of supergravity,
might provide us with a window on some interesting new corners of M-theory
moduli space.

The outline of this letter is as follows. We give a brief review of the field
equations of third order Lovelock gravity in Sec. \ref{Fiel}. In Sec. \ref
{8dNUT}, we obtain Taub-NUT solutions of third order Lovelock gravity in
eight dimensions and then we check the conjectures given in Ref. \cite
{DehMan}. We finish this letter with some concluding remarks.

\section{Field Equations\label{Fiel}}

The vacuum gravitational field equations of third order Lovelock gravity may
be written as:
\begin{equation}
\alpha _{0}g_{\mu \nu }+\alpha _{1}G_{\mu \nu }^{(1)}+\alpha _{2}G_{\mu \nu
}^{(2)}+\alpha _{3}G_{\mu \nu }^{(3)}=0,  \label{Geq}
\end{equation}
where $\alpha _{i}$'s are Lovelock coefficients, $G_{\mu \nu }^{(1)}$ is
just the Einstein tensor, and $G_{\mu \nu }^{(2)}$ and $G_{\mu \nu }^{(3)}$
are the second and third order Lovelock tensors given as
\begin{equation}
G_{\mu \nu }^{(2)}=2(R_{\mu \sigma \kappa \tau }R_{\nu }^{\phantom{\nu}%
\sigma \kappa \tau }-2R_{\mu \rho \nu \sigma }R^{\rho \sigma }-2R_{\mu
\sigma }R_{\phantom{\sigma}\nu }^{\sigma }+RR_{\mu \nu })-\frac{1}{2}%
\mathcal{L}_{2}g_{\mu \nu },  \label{Love2}
\end{equation}
\begin{eqnarray}
G_{\mu \nu }^{(3)} &=&-3(4R^{\tau \rho \sigma \kappa }R_{\sigma \kappa
\lambda \rho }R_{\phantom{\lambda }{\nu \tau \mu}}^{\lambda }-8R_{%
\phantom{\tau \rho}{\lambda \sigma}}^{\tau \rho }R_{\phantom{\sigma
\kappa}{\tau \mu}}^{\sigma \kappa }R_{\phantom{\lambda }{\nu \rho \kappa}%
}^{\lambda }+2R_{\nu }^{\phantom{\nu}{\tau \sigma \kappa}}R_{\sigma \kappa
\lambda \rho }R_{\phantom{\lambda \rho}{\tau \mu}}^{\lambda \rho }  \nonumber
\\
&&-R^{\tau \rho \sigma \kappa }R_{\sigma \kappa \tau \rho }R_{\nu \mu }+8R_{%
\phantom{\tau}{\nu \sigma \rho}}^{\tau }R_{\phantom{\sigma \kappa}{\tau \mu}%
}^{\sigma \kappa }R_{\phantom{\rho}\kappa }^{\rho }+8R_{\phantom
{\sigma}{\nu \tau \kappa}}^{\sigma }R_{\phantom {\tau \rho}{\sigma \mu}%
}^{\tau \rho }R_{\phantom{\kappa}{\rho}}^{\kappa }  \nonumber \\
&&+4R_{\nu }^{\phantom{\nu}{\tau \sigma \kappa}}R_{\sigma \kappa \mu \rho
}R_{\phantom{\rho}{\tau}}^{\rho }-4R_{\nu }^{\phantom{\nu}{\tau \sigma
\kappa }}R_{\sigma \kappa \tau \rho }R_{\phantom{\rho}{\mu}}^{\rho
}+4R^{\tau \rho \sigma \kappa }R_{\sigma \kappa \tau \mu }R_{\nu \rho
}+2RR_{\nu }^{\phantom{\nu}{\kappa \tau \rho}}R_{\tau \rho \kappa \mu }
\nonumber \\
&&+8R_{\phantom{\tau}{\nu \mu \rho }}^{\tau }R_{\phantom{\rho}{\sigma}%
}^{\rho }R_{\phantom{\sigma}{\tau}}^{\sigma }-8R_{\phantom{\sigma}{\nu \tau
\rho }}^{\sigma }R_{\phantom{\tau}{\sigma}}^{\tau }R_{\mu }^{\rho }-8R_{%
\phantom{\tau }{\sigma \mu}}^{\tau \rho }R_{\phantom{\sigma}{\tau }}^{\sigma
}R_{\nu \rho }-4RR_{\phantom{\tau}{\nu \mu \rho }}^{\tau }R_{\phantom{\rho}%
\tau }^{\rho }  \nonumber \\
&&+4R^{\tau \rho }R_{\rho \tau }R_{\nu \mu }-8R_{\phantom{\tau}{\nu}}^{\tau
}R_{\tau \rho }R_{\phantom{\rho}{\mu}}^{\rho }+4RR_{\nu \rho }R_{%
\phantom{\rho}{\mu }}^{\rho }-R^{2}R_{\nu \mu })-\frac{1}{2}\mathcal{L}%
_{3}g_{\mu \nu }.  \label{Love3}
\end{eqnarray}
In Eqs. (\ref{Love2}) and (\ref{Love3}) $\mathcal{L}_{2}=R_{\mu \nu \gamma
\delta }R^{\mu \nu \gamma \delta }-4R_{\mu \nu }R^{\mu \nu }+R^{2}$ is the
Gauss-Bonnet Lagrangian and
\begin{eqnarray}
\mathcal{L}_{3} &=&2R^{\mu \nu \sigma \kappa }R_{\sigma \kappa \rho \tau }R_{%
\phantom{\rho \tau }{\mu \nu }}^{\rho \tau }+8R_{\phantom{\mu \nu}{\sigma
\rho}}^{\mu \nu }R_{\phantom {\sigma \kappa} {\nu \tau}}^{\sigma \kappa }R_{%
\phantom{\rho \tau}{ \mu \kappa}}^{\rho \tau }+24R^{\mu \nu \sigma \kappa
}R_{\sigma \kappa \nu \rho }R_{\phantom{\rho}{\mu}}^{\rho }  \nonumber \\
&&+3RR^{\mu \nu \sigma \kappa }R_{\sigma \kappa \mu \nu }+24R^{\mu \nu
\sigma \kappa }R_{\sigma \mu }R_{\kappa \nu }+16R^{\mu \nu }R_{\nu \sigma
}R_{\phantom{\sigma}{\mu}}^{\sigma }-12RR^{\mu \nu }R_{\mu \nu }+R^{3}
\label{L3}
\end{eqnarray}
is the third order Lovelock Lagrangian. Equation (\ref{Geq}) does not
contain the derivative of the curvatures, and therefore the derivatives of
the metric higher than two do not appear. In order to have the contribution
of all the above terms in the field equation, the dimension of the spacetime
should be equal or larger than seven. Here, for simplicity, we consider the
NUT solutions of the dimensionally continued gravity in eight dimensions.
The dimensionally continued gravity in $D$ dimensions is a special class of
the Lovelock gravity, in which the Lovelock coefficients are reduced to two
by embedding the Lorentz group $SO(D-1,1)$ into a larger AdS group $%
SO(D-1,2) $ \cite{Zan}. By choosing suitable unit, the remaining two
fundamental constants can be reduced to one fundamental constant $l$. Thus,
the Lovelock coefficients $\alpha _{i}$'s can be written as
\begin{equation*}
\alpha _{0}=-\frac{21}{l^{2}},\text{ \ \ }\alpha _{1}=3,\text{ \ \ }\alpha
_{2}=\frac{3l^{2}}{20},\text{ \ \ \ }\alpha _{3}=\frac{l^{4}}{120}.
\end{equation*}

\section{Eight-dimensional Solutions\label{8dNUT}}
In this section we study the eight-dimensional Taub-NUT solutions of third
order Lovelock gravity. In constructing these metrics the idea is to regard
the Taub-NUT spacetime as a $U(1)$ fibration over a $6$-dimensional base
space endowed with an Einstein-K\"{a}hler metric $d\Omega {_{B}}^{2}$. Then
the Euclidean section of the $8$-dimensional Taub-NUT spacetime can be
written as:
\begin{equation}
ds^{2}=F(r)(d\tau +N\mathcal{A})^{2}+F^{-1}(r)dr^{2}+(r^{2}-N^{2})d\Omega {_{B}}^{2},
\label{TN}
\end{equation}
where $\tau $ is the coordinate on the fibre $S^{1}$ and $A$ has a curvature
$F=dA$, which is proportional to some covariantly constant 2-form. Here $N$
is the NUT charge and $F(r)$ is a function of $r$. The solution will
describe a `NUT' if the fixed point set of the $U(1)$ isometry $\partial
/\partial \tau $ (i.e. the points where $F(r)=0$) is less than $6$%
-dimensional and a `bolt' if the fixed point set is $6$-dimensional. Here,
we consider only the cases where all the factor spaces of $\mathcal{B}$ have
zero or positive curvature. Thus, the base space $\mathcal{B}$ can be the
6-dimensional space $\Bbb{CP}^{3}$, a product of three 2-dimensional spaces (%
$T^{2}$ or $S^{2}$), or the product of a 4-dimensional space $\Bbb{CP}^{2}$ with a 2-dimensional one. 
The $1$-forms and metrics of $S^{2}$, $T^{2}$, $\Bbb{CP}^{2}$ and $%
\Bbb{CP}^{3}$ are \cite{Pop}
\begin{eqnarray}
&& \mathcal{A}_{S^{2}}=2\cos \theta _{i}d\phi _{i},  \hspace{.5cm}
d\Omega _{S^{2}}^{2} =d\theta _{i}^{2}+\sin ^{2}\theta _{i}d\phi _{i}^{2},
\label{S2}\\
&& \mathcal{A}_{T^{2}} =2\eta _{i}d\zeta _{i},  \hspace{1.2cm}
d\Omega _{T^{2}}^{2} = d\eta _{i}^{2}+d\zeta _{i}^{2},  \label{T2}\\
&& \mathcal{A}_{\Bbb{CP}^{2}} =6\sin ^{2}\theta_{2}(d\phi _{2}+\sin ^{2}\theta_{1}d\phi
_{1}),
\label{ACP2} \\
&& d\Omega ^{2}{_{\Bbb{CP}^{2}}} =6\{d{\theta _{2}}^{2}+\sin ^{2}\theta _{2}\cos ^{2}\theta
_{2}(d\phi _{2}+\sin ^{2}\theta _{1}d\phi _{1})^{2}  \nonumber \\
&&\hspace{1.7cm} +sin^{2}\theta _{2}({d\theta _{1}}^{2}+\sin ^{2}\theta _{1}\cos ^{2}\theta _{1}%
{d\phi _{1}}^{2})\},  \label{CP2}\\
&& \mathcal{A}_{\Bbb{CP}^{3}} =\frac{1}{2}\left( \frac{1}{2}(\cos ^{2}\theta
_{3}-\sin ^{2}\theta _{3})d\phi _{3}-\cos ^{2}\theta _{3}\cos \theta
_{1}d\phi _{1}-\sin ^{2}\theta _{3}\cos \theta _{2}d\phi _{2}\right) ,
\nonumber \\
&& d\Omega ^{2}{_{\Bbb{CP}^{3}}} =8\{d\theta _{3}^{2}+\frac{1}{4}\sin
^{2}\theta _{3}\cos ^{2}\theta _{3}(d\phi _{3}-\cos \theta _{1}d\phi
_{1}+\cos \theta _{2}d\phi _{2})^{2}  \nonumber \\
&&\hspace{1.7cm}+\frac{1}{4}\cos ^{2}\theta _{3}(d\theta _{1}^{2}+\sin ^{2}\theta _{1}d\phi
_{1}^{2})+\frac{1}{4}\sin ^{2}\theta _{3}(d\theta _{2}^{2}+\sin ^{2}\theta
_{2}d\phi _{2}^{2})\},  \label{CP3}
\end{eqnarray}
respectively. To find the metric function $F(r)$, one may use any components
of Eq. (\ref{Geq}). After some calculation, we find that the metric function $F(r)$ for any base space
$\mathcal{B}$ can be written as

\begin{equation}
F(r)=\frac{1}{\Gamma l^{2}}\left( \frac{\Psi ^{1/3}}{2\beta }-\frac{3\left(
r^{2}-N^{2}\right) ^{4}B+\left( r^{2}-N^{2}\right) ^{2}l^{4}E}{3\Psi ^{1/3}}%
+\Omega \right),  \label{F(r)}
\end{equation}
where
\begin{eqnarray*}
\Psi &=&\left( r^{2}-N^{2}\right) \left( 4C+\sqrt{16C^{2}+8\beta ^{3}\left(
r^{2}-N^{2}\right) ^{4}\left[ \left( r^{2}-N^{2}\right) ^{6}B^{3}+\frac{%
l^{12}}{27}E^{3}\right] }\right) , \\
\Omega &=&\frac{p}{6}\left( r^{4}-N^{4}\right) l^{2}+\left(
r^{2}-N^{2}\right) \left( 5r^{2}+3N^{2}\right) , \\
\Gamma &=&5r^{4}+6N^{2}r^{2}+5N^{4},
\end{eqnarray*}
and $p$\ is the dimension of the curved factor spaces of $\mathcal{B}$. The
constant $\beta $ and the functions $B$, $E$\ and $C$ depend on the choice of
the base space as:\bigskip
\begin{center}
\begin{tabular}{llll}
\hline\hline
$\mathcal{B}$ &\  $\beta $ & \ $E$ & $B$ \\ \hline
${{\Bbb{CP}^{3}}}$ &\  $2$ & \ $0$ & $\left( 8N^{2}-l^{2}\right) ^{2}$ \\
$S^{2}\times \Bbb{CP}^{2}$ &\  $9$ & $\ 2\left(
13r^{4}-6r^{2}N^{2}+13N^{4}\right) $ & $72N^{2}\left( 4N^{2}-l^{2}\right) $
\\
$S^{2}\times S^{2}\times S^{2}$ & \ $3$ & \ $12\left( r^{4}+N^{4}\right) $ & $%
24N^{2}\left( 4N^{2}-l^{2}\right) $ \\
$T^{2}\times \Bbb{CP}^{2}$ & \ $9$ & $\ -4\left( r^{4}+6r^{2}N^{2}+N^{4}\right) $
& $48N^{2}\left( 6N^{2}-l^{2}\right) $ \\
$T^{2}\times S^{2}\times S^{2}$ & \ $3$ & \ $0$ & $\frac{2}{3}\left(
12N^{2}-l^{2}\right) ^{2}$ \\
$T^{2}\times T^{2}\times S^{2}$ & \ $3$ & \ $-2\left( r^{2}+N^{2}\right) ^{2}$ &
$8N^{2}\left( 12N^{2}-l^{2}\right) $ \\
$T^{2}\times T^{2}\times T^{2}$ & \ $1$ & \ $0$ & $32N^{4}$ \\ \hline\hline
\end{tabular}\\
\end{center}

\noindent and\\

\begin{tabular}{ll}
\hline\hline
$\mathcal{B}$ & $C$ \\ \hline
$
\begin{array}{c}
\Bbb{CP}{{^{3}}}
\end{array}
$ & $\left( l^{2}-8N^{2}\right) ^{3}\left( r^{2}+N^{2}\right) \Upsilon
+mr\Gamma ^{2}$ \\
$
\begin{array}{c}
S^{2}\times \Bbb{CP}^{2} \\
\text{ \ \ \ }
\end{array}
$ & $
\begin{array}{c}
324\left( r^{2}+N^{2}\right) \{18N^{4}\left( 3l^{2}-8N^{2}\right) \Upsilon
+l^{6}\left[ 7\left( r^{8}+N^{8}\right) +38r^{2}N^{2}\left(
N^{4}+r^{2}N^{2}+r^{4}\right) \right]  \\
-l^{4}N^{2}(93N^{8}+652r^{2}N^{6}+558r^{4}N^{4}+652r^{6}N^{2}+93r^{8})%
\}+81mr\Gamma ^{2}
\end{array}
$ \\
$
\begin{array}{c}
S^{2}\times S^{2}\times S^{2} \\
\text{ \ \ }
\end{array}
$ & $
\begin{array}{c}
72\left( r^{2}+N^{2}\right) \{3N^{4}\left( 3l^{2}-8N^{2}\right) \Upsilon
+l^{6}\left( 2N^{8}+9r^{2}N^{6}+10r^{4}N^{4}+9r^{6}N^{2}+2r^{8}\right)  \\
-6l^{4}N^{2}\left( 3N^{4}+2r^{2}N^{2}+3r^{4}\right) \left(
N^{4}+6r^{2}N^{2}+3r^{4}\right) \}+9mr\Gamma ^{2}
\end{array}
$ \\
$
\begin{array}{c}
T^{2}\times \Bbb{CP}^{2} \\
\text{ \ \ }
\end{array}
$ & $
\begin{array}{c}
108\left( r^{2}+N^{2}\right) \{108N^{4}\left( l^{2}-4N^{2}\right) \Upsilon
-l^{6}\left( r^{2}-N^{2}\right) ^{4}-6l^{4}N^{2}(9N^{8}+92r^{2}N^{6} \\
+54r^{4}N^{4}+92r^{6}N^{2}+9r^{8})\}+81mr\Gamma ^{2}
\end{array}
$ \\
$
\begin{array}{c}
T^{2}\times S^{2}\times S^{2} \\
\text{ \ \ }
\end{array}
$ & $
\begin{array}{c}
-2\left( r^{2}+N^{2}\right) \{18N^{2}\left(
l^{4}-12l^{2}N^{2}+48N^{4}\right) \Upsilon +l^{6}(r^{2}+N^{2})\left(
r^{2}-N^{2}\right) ^{2}(7N^{4} \\
+2r^{2}N^{2}+7r^{4})\}+9mr\Gamma ^{2}
\end{array}
$ \\
${\,}T^{2}\times T^{2}\times S^{2}$ & $4\left( r^{2}+N^{2}\right) \left[
108N^{4}\left( l^{2}-8N^{2}\right) \Upsilon +l^{6}\left( r^{4}-N^{4}\right)
^{2}+18l^{4}N^{2}(r^{2}-N^{2})^{4}\right] +27mr\Gamma ^{2}$ \\
${\,}
\begin{array}{c}
T^{2}\times T^{2}\times T^{2}
\end{array}
$ & $-64N^{6}\left( r^{2}+N^{2}\right) \Upsilon +mr\Gamma ^{2}$ \\
\hline\hline
\end{tabular}\\
\\
\noindent where
\begin{equation*}
\Upsilon =(11r^{8}+84N^{2}r^{6}+66N^{4}r^{4}+84N^{6}r^{2}+11N^{8}).
\end{equation*}
Although we have written the metric function $F(r)$ for specific values of
Lovelock coefficients belonging to dimensionally continued Lovelock gravity
in $8$ dimensions, the form of $F(r)$ for arbitrary values of $\alpha_i$'s is the
same as Eq. (\ref{F(r)}) with more complicated $E$, $B$ and $C$.

\subsection{Taub-NUT Solutions:}
The solutions given in Eq. (\ref{F(r)}) describe NUT solutions, if (i) $%
F(r=N)=0$ and (ii) $F^{\prime }(r=N)=1/(4N)$. The first condition comes from
the fact that all the extra dimensions should collapse to zero at the fixed
point set of $\partial /\partial \tau $, and the second one is to avoid
conical singularity with a smoothly closed fiber at $r=N$. Using these
conditions, one finds that the third order Lovelock gravity, in eight
dimensions admits NUT solutions only with $\Bbb{CP}^{3}$ base space when the
mass parameter is fixed to be
\begin{equation}
m_{n}=2N\left( 8N^{2}-l^{2}\right) ^{3}.
\end{equation}

Computation of the Kretschmann scalar at $r=N$ for the solutions in eight
dimensions shows that the spacetimes with base spaces $S^{2}\times
S^{2}\times T^{2}$, $S^{2}\times S^{2}\times S^{2}$, $T^{2}\times CP^{2}$ or
$S^{2}\times CP^{2}$ have a curvature singularity at $r=N$ in Einstein
gravity, while the spacetime with $\mathcal{B}=\Bbb{CP}^{3}$ has no
curvature singularity at $r=N$. Thus, the conjecture given in \cite{DehMan}
is confirmed for third order Lovelock gravity too. Here we generalize this
conjecture for Lovelock gravity as ``\textit{For the non-extremal NUT
solutions of Einstein gravity which have no curvature singularity at $r=N$,
the Lovelock gravity admits NUT solutions, while the Lovelock gravity does
not admit non-extremal NUT when the spacetime has curvature singularity at $%
r=N$}.'' Indeed, we have non-extreme NUT solutions in $2+2k$
dimensions with non-trivial fibration when the $2k$-dimensional base space
is chosen to be $\Bbb{CP}^{k}$. Although we have not written the solutions
in $2+2k$ dimensions and with arbitrary $\alpha_i$'s, but calculations confirm the above conjecture.

\subsection{Extreme Taub-NUT Solution}

The solutions (\ref{F(r)}) with the base space $\mathcal{B}_{2}=T^{2}\times
T^{2}\times T^{2}$ and $\mathcal{B}_{3}=S^{2}\times T^{2}\times T^{2}$
satisfy the extremal NUT solutions provided the mass parameter is fixed to
be
\begin{eqnarray}
m_{n}^{B_{2}} &=&128N^{7},  \label{MnTTT} \\
m_{n}^{B_{3}} &=&16N^{5}\left( 8N^{2}-l^{2}\right) .
\end{eqnarray}
Indeed for these two cases $F^{\prime }(r=N)=0$, and therefore the NUT
solutions should be regarded as extremal solutions. Computing the
Kretschmann scalar, we find that there is a curvature singularity at $r=N$
for the spacetime with $\mathcal{B}_{3}=S^{2}\times T^{2}\times T^{2}$,
while the spacetime with $\mathcal{B}_{2}=T^{2}\times T^{2}\times T^{2}$ has
no curvature singularity at $r=N$. This leads us to the generalization of
second conjecture of Ref. \cite{DehMan}: ``\textit{\ Lovelock gravity has
extremal NUT solutions whenever the base space is a product of 2-torii with
at most one }$2$\textit{-dimensional space of positive curvature}''. Indeed,
calculations in other dimensions show that when the base space has at most
one two dimensional curved space as one of its factor spaces, then Lovelock
gravity admits an extreme NUT solution even though there exists a curvature
singularity at $r=N$.

\section{Concluding Remarks \label{con}}

We considered the existence of Taub-NUT solutions in $8$-dimensional third
order Lovelock gravity with cosmological constant. Although one can do the
calculations for any arbitrary values of Lovelock coefficients, we chose
them as those of dimensionally continued Lovelock gravity in eight
dimensions to have more compact form for the solutions. It is worthwhile to
mention that this choice of Lovelock coefficients has no effect on the
properties if the solutions. These solutions are constructed as circle
fibrations over even dimensional spaces that in general are products of
Einstein-K\"{a}hler spaces. We found that as in the case of Gauss-Bonnet
gravity, the function $F(r)$ of the metric depends on the specific form of
the base factors on which one constructs the circle fibration. In other
words we found that the solutions are sensitive to the geometry of the base
space, in contrast to Einstein gravity where the metric in any dimension is
independent of the specific form of the base factors. Thus, one may say that
the sensitivity of the NUT solutions on the geometry of the base space is a
common feature of higher order Lovelock gravity, which does not happen in Einstein gravity.

We have found that when Einstein gravity admits non-extremal NUT
solutions with no curvature singularity at $r=N$, then there exists a
non-extremal NUT solution in third order Lovelock gravity. In $8$%
-dimensional spacetime, this happens when the metric of the base space is
chosen to be $\Bbb{CP}^{3}$. Indeed, third order Lovelock gravity does not
admit non-extreme NUT solutions with any other base spaces. Although we have not written the NUT solutions in
other dimensions, we found that in any dimension $2k+2$, we have only one
non-extremal NUT solution with $\Bbb{CP}^{k}$ as the base space. That is, the
Lovelock gravity singles out the preferred non-singular base space $\Bbb{CP}%
^{k}$ in $2+2k$ dimensions. We also
found that only when the base space is $T^{2}\times T^{2}\times T^{2}$ or $%
S^{2}\times T^{2}\times T^{2}$, eight-dimensional third order Lovelock
gravity admits extremal NUT solution. Calculations show that the extremal
NUT black holes exist for the base spaces $T^{2}\times T^{2}\times
T^{2}.....\times T^{2}$ and $S^{2}\times T^{2}\times T^{2}.....\times T^{2}$.
We have extended these observations to two conjectures about the existence
of NUT solutions in Lovelock gravity. The study of thermodynamic properties
of these solutions remains to be carried out in future.

\acknowledgments{ This work has been supported financially by
Research Institute for Astronomy and Astrophysics of Maragha.}

\end{document}